\documentclass[
    ,final            
  ]
  {aipproc}

\layoutstyle{6x9}

\newcommand{\bm}[1]{\mbox{\boldmath $#1$}}

\newcommand{\open}{{<\kern -0.3 em{\scriptscriptstyle )}}}

\newcommand{\nslash}{\kern 0.2 em n\kern -0.45em /}
\newcommand{\Pslash}{\kern 0.2 em P\kern -0.56em \raisebox{0.3ex}{/}}
\newcommand{\pslash}{\kern 0.2 em p\kern -0.4em /}
\newcommand{\kslash}{\kern 0.2 em k\kern -0.45em /}
\newcommand{\Sslash}{\kern 0.2 em S\kern -0.56em \raisebox{0.3ex}{/}}

\newcommand{\eq}{\begin{equation}}
\newcommand{\ee}{\end{equation}}
\newcommand{\beq}{\begin{equation}}
\newcommand{\eeq}{\end{equation}}
\newcommand{\ba}{\begin{eqnarray}}
\newcommand{\ea}{\end{eqnarray}}
\newcommand{\eqa}{\begin{eqnarray}}
\newcommand{\eea}{\end{eqnarray}}

\newcommand{\sumint}{\kern 0.2 em {\textstyle\sum} \kern -1.1 em \int}

\begin{document}

\title{Asymmetric parton distributions of the nucleon}


\author{Dani\"el Boer}{
  address={KVI, University of Groningen, Zernikelaan 25, NL-9747 AA Groningen, The Netherlands}
}

\begin{abstract}
This contribution to CIPANP 2012 highlights what we have learned about the asymmetric parton distributions of the nucleon over the past 20 years. 
These distributions include the transverse momentum dependent parton distributions describing spin-orbit correlations, but also their generalized 
parton and Wigner distribution analogues. Besides quark distributions, also gluon distributions are discussed, in particular the distribution of 
linearly polarized gluons inside unpolarized nucleons and its applications, such as at LHC in Higgs production and at a future Electron-Ion Collider 
in heavy quark production. 
\end{abstract}

\maketitle


\section{Asymmetric spin structure of nucleons}
When viewing a proton with a probe that has high enough energy to resolve the partonic structure, 
it matters whether the proton has its spin directed towards the probe or orthogonal to it. In the 
latter case the distributions of quarks probed inside the proton (here assumed to be at rest for simplicity) 
can be left-right 
asymmetric w.r.t.\ the plane spanned by the probe momentum and the transverse spin direction.
These asymmetric distributions can result in asymmetric production of final state particles in collisions with transversely polarized protons, 
which has most clearly been observed in the process $p^\uparrow p \to \pi\, X$, where the pion transverse momentum ($p_T$) distribution is
highly left-right asymmetric  
\cite{Adams:1991cs,Adams:1991ru,Bravar:1996ki,Krueger:1998hz,Adams:2003fx,Abramov:2005ax,Arsene:2008aa,Abelev:2008af}. 
In fixed target experiments at large $x_F=2p_z/\sqrt{s}$, 
the asymmetry reaches up to 40\% for charged pions. 
Such a single spin asymmetry is actually a $\sin (\phi_S)$ distribution \cite{Poljak:2009pd}, where $\phi_S$ is the azimuthal angle of the transverse spin vector $S_T$. This $p_T \times S_T$ correlation is indicative of an underlying spin-orbit coupling and persists out to high energies (at RHIC 
$\sqrt{s}=200$ GeV). At such high energies a factorized description is expected to be applicable. Since any asymmetry 
generated in the hard partonic scattering is tiny \cite{Kane:1978nd}, it must have a nonperturbative origin. 
In \cite{Sivers:1989cc,Sivers:1990fh} D. Sivers put forward the 
suggestion that there is an asymmetry in the quark transverse momentum ($k_T$) distribution in a transversely polarized proton, a 
nonperturbative $k_T\times S_T$ correlation.  
Although such a Sivers effect can effectively describe the asymmetry data \cite{Anselmino:1994tv,Anselmino:1999pw}, its theoretical definition as a 
transverse momentum dependent parton distribution (TMD) is not straightforward and has been modified several times 
over the years \cite{Boer:1997nt,Collins:2002kn,Belitsky:2002sm,Ji:2004wu,Ji:2004xq,Collins:2011zzd}.
The definition of TMDs is closely tied to the factorization of the scattering processes considered. Also TMD factorization proofs have 
been modified and completed over the years \cite{Collins:1981uk,Ji:2004wu,Ji:2004xq,Collins:2011zzd}, 
resulting very recently in a TMD factorization proof \cite{Collins:2011zzd} that resolves "several difficulties and inconsistencies of earlier versions" as briefly summarized in \cite{Collins:2011ca}. This factorization applies specifically to the processes of semi-inclusive 
DIS (${e \, p \to e' \, {h} \, X}$) and the Drell-Yan process (${p \, p \to \mu \bar{\mu} \, X}$), 
but {\em not\/} to $p \, p \to \pi \,X$. At large $p_T$, $p^\uparrow p \to \pi\, X$ can be described by collinear factorization beyond leading twist, which will not be discussed here. Semi-inclusive DIS and Drell-Yan involve
electrons or muons, i.e.\ virtual photons, which provide for cleaner observables directly sensitive to the TMD effects of interest. 
According to the TMD formalism the Sivers effect in semi-inclusive DIS should show up as a $\sin(\phi_h-\phi_S)$ asymmetry \cite{Boer:1997nt}, 
which has been clearly observed by the HERMES \cite{Airapetian:2009ae} and COMPASS \cite{Alekseev:2010rw} experiments. 
The Sivers effect is also going to be studied extensively in Drell-Yan, where it should lead to 
the opposite sign for the asymmetry \cite{Collins:2002kn}. 
The reason for this difference is that the theoretical definition of the Sivers TMD $f_{1T}^{\perp}$ is not unique:
\beq  
P \! \cdot \! \left({\bf k}_T \times {\bf S}_T \right) {f_{1T}^{\perp [{\cal C}]}(x,{{\bf k}_T^2})} \propto\ {\rm F.T.} \, \langle P, {S_T}| \bar{\psi}(0) \, {\cal L}_{{\cal C}[0,\xi]} \, \gamma^+ \, \psi(\xi) |P, {S_T}\rangle\big|_{\xi = (\xi^-,0^+,{\bf \xi}_T)}, 
\label{Siversdef}
\eeq
where F.T. stands for Fourier transform and ${\cal L}_{{\cal C}[0,\xi] }$ for the Wilson line 
\beq 
{\cal L}_{{\cal C}[0,\xi]} = {\cal P} \exp \left(-ig\int_{{{\cal C}[0,\xi]}} 
ds_\mu \, A^\mu (s)\right), 
\eeq 
whose path ${\cal C}$ from $0$ to $\xi$ turns out to depend on the process! This does not mean that predictability is lost however, 
since it is a {\em calculable\/} process dependence.
Calculation shows that the proper gauge invariant definition of TMDs in semi-inclusive DIS 
contains a future pointing Wilson line arising due to final state interactions (FSI), whereas in Drell-Yan (DY) it is 
past pointing due to initial state interactions (ISI) \cite{Brodsky:2002cx,Brodsky:2002rv,Collins:2002kn,Belitsky:2002sm}. 
Time reversal invariance relates the Sivers functions of SIDIS and Drell-Yan, which yields the sign relation 
$ f_{1T}^{\perp [{\rm SIDIS}]} = - f_{1T}^{\perp [{\rm DY}]}$ \cite{Collins:2002kn} that remains to be tested. It is a prediction that follows from TMD factorization.
In processes where partons in more than two hadrons play a role, one has to take into account both ISI and FSI \cite{Bomhof:2004aw}, 
leading to more complicated paths of the Wilson lines. It can even lead to entangled Wilson lines that cannot be factorized.  
TMD factorization fails for processes like $p\, p\to  h_1 \, h_2\, X$ \cite{Collins:2007nk,Collins:2007jp,Rogers:2010dm}, 
except for certain projections of it, i.e.\ particular transverse momentum weighted asymmetries (an example is given in \cite{Bacchetta:2005rm}). 
It is important to stress that this factorization breaking does not cast doubt on the above sign relation.

The Sivers function {\em at tree level\/} involves a Wilson line that goes along the lightcone to infinity. Because this lightlike line remains after taking Mellin moments 
($\int dx\, x^{n}\, f_{1T}^{\perp  [{\cal C}]}(x,{{\bf k}_T^2})$), it is not amenable to lattice calculations using Euclidean space. 
However, beyond tree level the Wilson 
line along the lightcone is to be viewed as the limit of infinite rapidity and infinite extent $L$ of the Wilson line. For finite rapidities and finite $L$
certain (still nonlocal) projections of the Sivers function {\em can\/} be evaluated on the lattice \cite{Musch:2011er}. More specifically, after taking Mellin moments {\em and\/} a Bessel weighting proposed 
recently in \cite{Boer:2011xd}, the Sivers TMD then yields a well-defined, calculable quantity $\langle k_T \times S_T\rangle (n,r_T)$ 
-the average transverse momentum shift orthogonal to a given transverse polarization-. Here the Fourier conjugate $r_T$ of $k_T$ indicates 
the transverse ``size'' of the quarks probed. Recently, this Sivers shift has been calculated on the lattice as a function of $L$, for the first Mellin moment ($n=1$) and $r_T$ roughly in the range 0.1-0.5 fm. It is found that the Sivers shift for up minus down quarks is zero at $L=0$ (as it should), becomes negative and saturates to a finite value for positive $L$ (SIDIS case) and similarly for negative $L$ (DY case) but with opposite sign values \cite{Musch:2011er}. This is the first `first-principles' demonstration that the Sivers function as given in Eq.\ (\ref{Siversdef}) is nonzero within QCD and it clearly corroborates the sign change relation!  

There are further asymmetric TMDs describing {\em polarized\/} quarks inside {\em polarized\/} hadrons, such as the ``wormgear'' functions $g_{1T}$ (longitudinally polarized quarks inside transversely polarized nucleons) and $h_{1L}^\perp$ (transversely polarized quarks inside longitudinally polarized nucleons), 
and the ``pretzelosity'' function $h_{1T}^\perp$ (transversely polarized quarks inside transversely polarized nucleons), but these 
will not be discussed here, cf.\ \cite{Ralston:1979ys,Kotzinian:1994dv,Tangerman:1994eh,Mulders:1995dh}. Next we turn to polarized quarks inside unpolarized hadrons.

\section{Asymmetric structure of spin averaged nucleons}

In the case of the spin averaged Drell-Yan process ${p \, p \to \mu \bar{\mu} \, X}$, the differential cross section can have several 
angular modulations:
\beq
\frac{1}{\sigma}\frac{d\sigma}{d\Omega} \propto \left( 1+ {\lambda} \cos^2\theta + {\mu} \sin 2\theta \, \cos\phi + \frac{{\nu}}{2} \sin^2 \theta \, {\cos 2\phi} \right) 
\eeq
To order $\alpha_s$ the Lam-Tung relation connects two of these terms: $1-{\lambda}-2{\nu}=0$. In experiments at CERN and Fermilab,
large deviations from the Lam-Tung relation were observed in pion-nucleon Drell-Yan \cite{Falciano:1986wk,Guanziroli:1987rp,Conway:1989fs}. 
This indicates a failure of the collinear perturbative QCD treatment: with collinear parton densities, only higher order gluon emission can generate deviations from the Lam-Tung relation, but the NNLO ${\cal O}(\alpha_s^2$) 
result  \cite{Brandenburg:1993cj,Mirkes:1994dp} is (at least) an order of magnitude smaller and of opposite sign! The large deviation can  
naturally be explained \cite{Boer:1999mm} by transverse polarization of quarks inside unpolarized protons \cite{Boer:1997nt}. This requires nonzero quark transverse 
momentum and is described by a TMD $h_1^\perp$, also referred to as Boer-Mulders function. A similar $\cos 2 \phi$ asymmetry has been studied in semi-inclusive DIS
by the HERMES and COMPASS experiments  \cite{Giordano:2009hi,Kafer:2008ud}, but due to the lower energy also power suppressed higher twist terms, referred to as the Cahn effect, can contribute considerably to this observable, hampering an unambiguous extraction \cite{Barone:2009hw}. 
Note that the overall sign of $h_1^\perp$ does not enter in the $\cos 2 \phi$ asymmetry in DY, hence the expected sign difference between SIDIS and DY cannot be tested using this observable.

Again after taking Mellin moments and appropriate Bessel weighting of this TMD, a well-defined quantity is obtained that can be calculated 
on the lattice: $\langle k_T \times s_T\rangle (n,r_T)$, the average transverse momentum shift orthogonal to a given transverse polarization of quarks inside 
an unpolarized proton. The first lattice calculation of this quantity clearly shows a nonzero "Boer-Mulders shift" \cite{Musch:2011er}. This means that indeed the transverse momentum distribution of quarks is asymmetric even when averaging over all proton spin directions! Most processes are not sensitive to this asymmetry, but clearly 
not all proton spin averaged processes involve automatically an average over quark spins. The ${\cos 2\phi}$ term in the Drell-Yan process is one example. 

Such an asymmetry of transversely polarized quarks inside unpolarized protons not only appears in the transverse momentum distribution, but also in the spatial distribution, to which we turn next. 

\section{Asymmetric spatial structure of nucleons}

The process of Deeply Virtual Compton Scattering (DVCS) $\gamma^* \, p \to \gamma \, p'$ allows to probe the spatial distribution of quarks inside nucleons. 
The theoretical description of DVCS involves Generalized Parton Distributions (GPDs), which are off-forward matrix elements ($p^\prime\neq p$). GPDs are functions of 
several variables, one of them, $b_\perp$, gives the transverse spatial distance of quarks w.r.t.\ the ``center'' of the proton, to be specific: the transverse center of longitudinal 
momentum  ${\bf R}_\perp^{CM} \equiv \sum_i x_i {\bf r}_{\perp i} $ \cite{Burkardt:2000za,Burkardt:2002hr,Soper:1976jc}. 
Here $b_\perp$ is the Fourier conjugate of the  
exchanged transverse momentum $\Delta_\perp = (p'-p)_\perp$. It is not to be confused with the Fourier conjugate of $k_T$ of the TMDs (earlier indicated by $r_T$), 
the transverse spatial size. Orthogonal vectors will be indicated by $\perp$, rather than $T$, because different frames can be considered for the GPDs and TMDs. However, in the infinite momentum frame (IMF) they coincide and GPDs as function of $x$ and $b_\perp$ (zero skewness) can be interpreted as densities \cite{Burkardt:2002hr}.  
The spin dependence of these densities have been studied with lattice QCD by the QCDSF and UKQCD collaboration \cite{Gockeler:2006zu}, showing that even pions have a nontrivial spin structure \cite{Brommel:2006zz,Brommel:2007xd}. 
It is found that both the $b_\perp \times S_\perp$ and $b_\perp \times s_\perp$ correlations in the spatial quark distributions are nonzero.   
The $b_\perp \times S_\perp$ correlation (analogue of the Sivers effect) for up and down quarks is of opposite sign, but the $b_\perp \times s_\perp$ correlation is of same sign. This agrees with model expectations \cite{Burkardt:2007xm}. For a model study of the spin structure of kaons, including strange quarks, cf.\ \cite{Nam:2011yw}.

A natural question to ask is whether there is a possible relation between the asymmetric TMDs and GPDs, i.e.\ 
\[
\langle b_\perp \times s_\perp \rangle  \stackrel{?}{\Leftrightarrow} \langle k_T \times s_T\rangle
\]
Intuitively one might expect a relation between an excess of left-movers over right-movers and an excess of particles on the left, they are after all confined to move in a restricted space. But to formulate such a relation on the operator level is not straightforward. 
GPDs and TMDs can be obtained from the same underlying Wigner-type distribution, but are not easily related for given $k_T$ and/or $b_\perp$. But it does mean that for a particular proton state (e.g.\ in a Fock state expansion) one can calculate both GPDs and TMDs with the same wave functions, using the same parameters. But the bottom line is that there is no (model independent) proof that if TMDs are asymmetric, GPDs have to be too, and vice versa. 
An `effective' relation between the Sivers TMD and the GPD $E(x,b_\perp)$ has been put forward in \cite{Burkardt:2003uw,Burkardt:2003je} by substituting for the operator $\hat{{\bf I}}$ in correlators of the form   
$\langle \bar \psi\; \hat{{\bf I}}\; \psi \rangle$, some average of it, yielding $\bar{\bf I}\; \langle \bar \psi\psi \rangle$, where the average $\bar{\bf I}(b_\perp^2)$ is called the lensing function. It can be obtained within models explicitly 
\cite{Burkardt:2003uw,Burkardt:2003je,Meissner:2007rx,Gamberg:2009uk}. The resulting relation is:
\beq
{f_{1T}^{\perp (1)}(x)} \equiv 
\int d^2 \bm{k}_T
{\frac{\bm{k}_T^2}{2M^2}} f_{1T}^{\perp } (x, \bm{k}_T^2) 
\propto S_T \times b_\perp
\int d b_\perp^2 \bar{\bf I}(b_\perp^2)
\frac{\partial}{\partial b_\perp^2} {E(x, b_\perp^2)}
\eeq
A similar relation holds for the asymmetric quark spin dependent TMD $h_1^\perp$ and the chiral-odd GPD combination $2\tilde{H}_T+E_T$ \cite{Burkardt:2005hp,Burkardt:2005jm}. Such relationships help to predict and verify signs of the distributions for the different flavors. 
But a model independent analysis of the lensing function will be hard to obtain.

As said, TMDs and GPDs derive from the same Wigner-type distribution, which is a function of the variables $x,k_T,b_\perp$. Although it can be defined, there is no 
experiment known that could probe this distribution directly. Interestingly, it displays additional asymmetries that are neither present in the TMDs nor in the GPDs. 
For instance, in unpolarized protons there can be distortions of the form $k_T \cdot b_\perp$ in the unpolarized quark distribution \cite{Lorce:2011kd}.
After integration over either $k_T$ or $b_\perp$ this effect averages to zero. 

One can define an even more general Wigner distribution, which also depends on the longitudinal spatial direction $z$. After integration over $x$ and $k_T$ this yields three-vector $\vec{r}=(b_\perp,z)$ dependent Fourier transforms of form factors \cite{Burkardt:2000za}. For a very brief explanation of Wigner distributions, GPDs, TMDs, choice of frames and interpretations, cf.\ \cite{Boer:2004ai}. For an overview of the distortions of transverse spatial ($b_\perp$) charge densities of polarized nucleons and deuterons, their relation to form factors and electric and magnetic multipole moments, cf.\ the talk by Marc Vanderhaeghen at CIPANP 2012 \cite{Vanderhaeghen2012}, or Refs.\ \cite{Carlson:2007xd,Carlson:2008zc,Lorce:2010sj}. For discussion and pictures of three-dimensional shapes of the nucleon, cf. \cite{Miller:2003sa,Kvinikhidze:2006ty}. 

\section{Asymmetric gluon distributions}

Gluon TMDs also can be asymmetric. There is the gluon Sivers function, which is the distribution of unpolarized gluons inside a transversely polarized nucleon, 
but since gluons have spin 1, the analogue of the transversely polarized quarks inside an unpolarized hadron is quite a different quantity. There can be 
gluon polarization inside unpolarized hadrons \cite{Mulders:2000sh}, 
but it refers to linear polarization, which is an interference between $+1$ and $-1$ helicity states. Although the distribution of linearly polarized gluons inside unpolarized protons is nowadays often
denoted by $h_1^{\perp\, g}$, it is chiral-even, T-even and $k_T$-even (rank 2), as opposed to $h_1^{\perp \, q}$.   
Using the two real linear polarization vectors $\varepsilon_x^i=(1,0)$ and $\varepsilon_y^i=(0,1)$ as a basis, 
unpolarized gluons (with distribution $f_1^g$ or simply $g$), circularly polarized gluons ($g_1^g$ or $\Delta g$), and linearly polarized 
gluons ($h_1^{\perp\, g}$), correspond to the following density matrices, respectively:
\[
\left( \begin{array}{cc} 1 & 0 \\ 0 & 1 \end{array} \right),\
\left( \begin{array}{cc} 0 & i \\ -i & 0 \end{array} \right),\ \left( \begin{array}{cc} \cos 2\phi  & \sin 2\phi \\ \sin2\phi & -\cos 2\phi \end{array} \right),
\]
where $\phi$ is the angle between $\varepsilon_x$ and $k_T$. It means that the linearly polarized gluons described by $h_1^{\perp \, g}$ prefer to be polarized along $k_T$,  
with a $\cos 2\phi$ distribution around it ($\varepsilon_{\max}^i=\hat k_T^i=(\cos\phi,\sin\phi)$ and $-\varepsilon_{\max}^i$ have the highest probability, whereas $\varepsilon_{\min}^i=(\sin\phi,-\cos\phi)$ and $-\varepsilon_{\min}^i$ the lowest). See also the related discussion in Ref.\ 
\cite{Dominguez:2011br}. 

Linearly polarized gluons are generated perturbatively \cite{Nadolsky:2007ba,Balazs:2007hr,Catani:2010pd}, but 
a nonperturbative distribution ($h_1^{\perp\, g}$) can be present too \cite{Mulders:2000sh}.
Currently no experimental information on this distribution is available. Several suggestions on how it can measured have been put forward
recently. It can be probed for instance in charm and bottom quark pair production \cite{Boer:2010zf}, where the transverse momenta 
$\vec{K}^Q_{\perp}$ and $\vec{K}^{\bar Q}_\perp$ of the quark and antiquark should be large, in such a way that their sum is much smaller in magnitude than their difference. Nonzero $h_1^{\perp\, g}$ leads for instance 
to a $\cos 2(\phi_T-\phi_\perp)$ asymmetry in heavy quark pair production in DIS, where $\phi_{T/\perp}$ denote the angles of $\vec{K}^Q_{\perp} \pm \vec{K}^{\bar Q}_\perp$. Although ISI or FSI are not required for this TMD to be nonzero, there is no reason to assume it is process independent. Extractions from $e\,p\to Q \bar{Q}\, X$ or from 
$p\,p \to Q \bar{Q}\, X$ may thus differ. In fact, TMD factorization is expected to be broken for the latter process \cite{Collins:2007nk,Collins:2007jp,Rogers:2010dm}, except perhaps for particular moments. Therefore, measuring $h_1^{\perp\, g}$ through open heavy quark pair 
production, is best done at a future Electron-Ion Collider \cite{Boer:2011fh} or LHeC \cite{AbelleiraFernandez:2012cc}. 

Linearly polarized gluons also enter Higgs production ($\sigma (Q_T)$) at NNLO pQCD \cite{Catani:2010pd}. 
The nonperturbative distribution $h_1^{\perp \, g}$ can be present at tree level and would affect Higgs production at low $Q_T$.
Higgs production happens mainly through $gg \to H$, in which the contribution from linear polarization enters for both gluons.
Although the LHC collides unpolarized protons and is often called a gluon collider, it means that moreover it is in principle 
a {\it polarized} gluon collider.

Unlike the heavy quark pair production case, linear polarization of gluons modifies the transverse momentum ($q_T$) distribution of Higgs 
production in an angular independent manner. It leads to a characteristic modulation as a function of $Q_T=|\vec{q}_T|$, with overall sign determined 
by the parity of the Higgs \cite{Boer:2011kf}. It thus offers a means to determine whether the Higgs boson 
is a scalar or a pseudoscalar, {\em if\/} the distribution $h_1^{\perp \, g}$ is sufficiently large at energies around the Higgs mass. 

In reality the Higgs boson decays very fast, which means that there will be background processes to deal with, which generally 
will dilute the modulation.
For example, in the $H\to \gamma \gamma$ channel, linearly polarized gluons also enter in the process $gg \to \gamma \gamma$ 
without Higgs \cite{Nadolsky:2007ba, Qiu:2011ai}. It is thus a channel to possibly measure $h_1^{\perp \, g}$, but perhaps not to 
determine the parity of the Higgs. Explicit calculation shows that the latter is discernible only in a narrow region around the Higgs 
mass, determined by the very small Higgs decay width \cite{Boer:2011kf}.
The experimental energy resolution $\Delta Q$ then becomes important, but assuming a realistic value of $\Delta Q=0.5$ GeV 
for the $\gamma \gamma$ channel at CMS or ATLAS, it is found that the characteristic differences between positive and negative parity 
Higgs boson production are not washed out completely \cite{denDunnen:2012ym}. 

\section{Concluding remarks}

This overview has addressed asymmetries in the transverse momentum and spatial distributions of partons inside the nucleon, in particular asymmetries w.r.t.\ the spin direction of either the nucleon or its constituents. Since asymmetric TMDs, such as the Sivers effect, display a process dependence, it is natural to wonder whether such distributions actually refer to properties of the proton? TMD factorization allows assigning the asymmetries to the nucleon while isolating the process dependence in numerical pre-factors, but the separation of effects in TMD factorization is not unique. Factorization separates long and short distance contributions, and can be organized such that long distance pieces can be associated with different hadrons, but that is in part a matter of choice. At the very least there is some ambiguity in assigning effects to the proton or to the partonic scattering process.
Here it can be useful to make a distinction between `static' and `dynamic' distributions, based on whether the quantity is calculable solely in terms of squared light-cone wave functions of the nucleon (probabilities) or whether it requires phases (Wilson lines) that depend on the initial and/or final state interactions in a process, cf.\ \cite{Brodsky:2002ue,Brodsky:2009dv}. Also dynamic distributions reflect properties of the proton, despite the process dependence, but the question is to what extent.  

The asymmetric {\it spatial} distributions are process independent and are generally considered properties of the proton. They give rise to electric and magnetic moments that can be measured at low energies. The transverse momentum asymmetries may manifest themselves also at lower energies, but one can never extract TMDs without a high-energy process. Therefore, they describe how the proton is seen at high energies. Different probes yielding different asymmetries may then simply be like looking at a nonsymmetric object in different ways, e.g.\ directly, via a mirror or through some filtering glasses. By looking with many different probes one may obtain a picture of what is the underlying momentum asymmetry in the nucleon and what is the effect of the probing processes themselves. How the transverse momentum and spatial asymmetries tie together precisely will hopefully be clarified further in the future. 


\begin{theacknowledgments}
I would like to thank the organizers for their
kind invitation to give a plenary talk at CIPANP 2012. Also, I thank my collaborators on several of the 
subjects reviewed here.   
\end{theacknowledgments}



\bibliographystyle{aipproc}   


\hyphenation{Post-Script Sprin-ger}

\IfFileExists{\jobname.bbl}{}
 {\typeout{}
  \typeout{******************************************}
  \typeout{** Please run "bibtex \jobname" to optain}
  \typeout{** the bibliography and then re-run LaTeX}
  \typeout{** twice to fix the references!}
  \typeout{******************************************}
  \typeout{}
 }

\end{document}